\documentclass[11pt]{article}

\usepackage[preprint]{acl}

\usepackage{times}
\usepackage{latexsym}

\usepackage[T1]{fontenc}

\usepackage[utf8]{inputenc}

\usepackage{microtype}

\usepackage{inconsolata}

\usepackage{graphicx}

\usepackage{amsmath}
\usepackage{amssymb}
\usepackage{algorithm}
\usepackage{algpseudocode}
\usepackage{xcolor}

\usepackage{graphicx}     
\usepackage{xcolor}       
\usepackage{fontawesome5} 
\usepackage{hyperref}     

\definecolor{IconRed}{HTML}{B2483F}
\usepackage[table]{xcolor}
\makeatletter
\newcommand{\authornote}[1]{%
  \ifcase#1\or \dagger\or \ddagger\or *\fi
}
\renewcommand{\@fnsymbol}[1]{\ensuremath{\authornote{#1}}}
\makeatother
\definecolor{DynaBlue}{HTML}{0B5CAD}
\definecolor{DynaNavy}{HTML}{05295C}
\definecolor{DynaLight}{HTML}{E9F2FC}
\usepackage{booktabs}
\usepackage{tabularx}
\usepackage{xcolor}
\usepackage{colortbl}

\definecolor{DynaTableRed}{HTML}{CC0000}
\definecolor{DynaTableYellow}{HTML}{FFBF2B}
\definecolor{DynaTableHeader}{HTML}{FFF1CC}
\definecolor{DynaTableHighlight}{HTML}{FBECEC}

\title{DynaFlow: Learning Free-Form Multi-Agent Orchestration}

\author{
  Ruhao Liu\thanks{Equal contribution.}
  \quad Qi Li\footnotemark[1]
  \quad Shizun Wang\thanks{Project Lead.}
  \quad Xinchao Wang\thanks{Corresponding Author.} \\
  National University of Singapore \\
  \texttt{\{ruhaoliu, liqi, shizun.wang\}@u.nus.edu}
  \quad \texttt{xinchao@nus.edu.sg}
  \\[0.5em]
  {
    \href{https://github.com/amiya-special/Dynaflow}{
      \textcolor{IconRed}{
        \raisebox{-0.1em}{\faGithub}
        \hspace{0.3em}Code
      }
    }
    \hspace{1.2em}
    \href{https://amiya-special.github.io/DynaFlow-website}{
      \textcolor{IconRed}{
        \raisebox{-0.1em}{\faGlobe}
        \hspace{0.3em}Project Page
      }
    }
  }
}
\begin{document}
\maketitle
\begin{abstract}
Multi-agent systems often rely on predefined collaboration structures, limiting their ability to adapt to diverse task requirements. We introduce \texttt{DynaFlow}, a framework that integrates free-form executable workflow synthesis into the action space of the main agent. The agent decides when collaboration is needed and generates task-specific programs that coordinate sub-agents through sequential or parallel execution, structured information exchange, and verification. Execution results and traces inform subsequent actions, enabling the agent to adapt collaboration as reasoning progresses. We train only the main agent through supervised fine-tuning on distilled trajectories, followed by refinement on examples filtered for unsupported premises, biased verification, and information-flow defects. All sub-agents remain fixed, separating orchestration learning from executor capability improvement. Across six benchmarks spanning mathematical reasoning, scientific question answering, and multi-hop retrieval, DynaFlow achieves 88.08\% overall accuracy in Forced mode, outperforming the strongest evaluated baseline by 10.90 percentage points with the same Qwen3.5-27B backbone. Workflow analysis finds more explicit use, verification, and refinement of upstream outputs after training, highlighting cross-stage information flow as a key dimension of multi-agent orchestration.
\end{abstract}

\begin{figure}[t]
  \centering
  \includegraphics[width=\columnwidth]{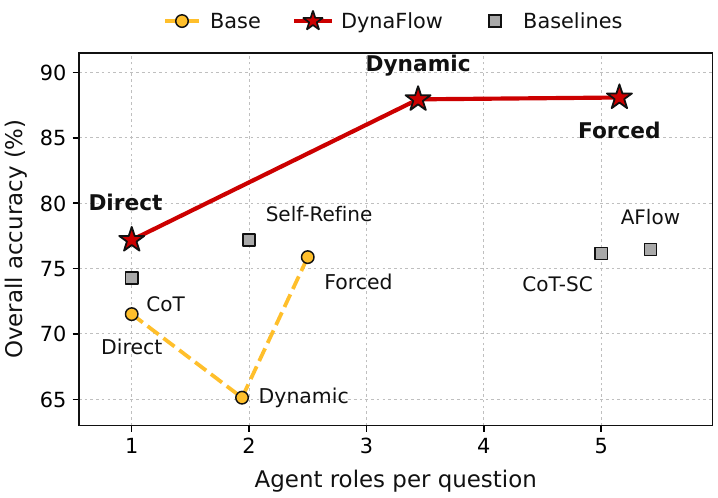}
\caption{\textbf{Overall accuracy versus collaboration size.}
Accuracy versus the mean number of agent roles per question on 688 questions (Qwen3.5-27B).  In Forced mode, DynaFlow improves accuracy from Base's 75.9\% to 88.1\%. Dynamic achieves 87.9\% with 3.44 agent roles per question on average, while answering 47\% of questions without workflows.}
  \vspace{-4mm}
  \label{fig:teaser}
\end{figure}

\begin{figure*}[t]
  \centering
  \includegraphics[width=\textwidth]{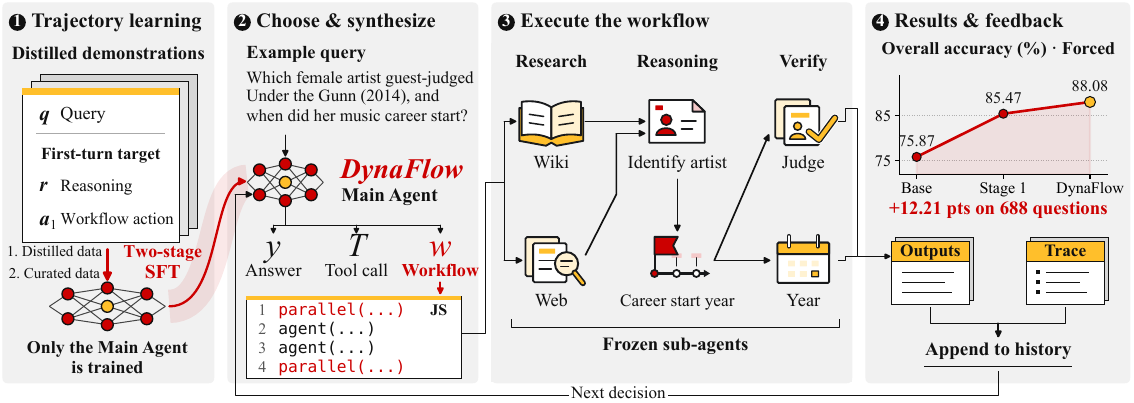}
\caption{\textbf{Overview of DynaFlow.}
(1) Two-stage SFT trains only the Main Agent on first-turn reasoning $r$ and workflow action $a_1$, using distilled demonstrations followed by a curated subset.
(2) It chooses to answer directly, call a tool, or generate a JavaScript workflow.
(3) Frozen sub-agents collaborate within the workflow according to the Main Agent's program.
(4) Outputs and execution traces update the history for subsequent decisions.}
  \label{fig:overview}
  \vspace{-4mm}
\end{figure*}

\section{Introduction}
\label{sec:introduction}

The effectiveness of multi-agent systems (MAS) built on large language models (LLMs) depends on how collaboration is organized. Many systems rely on fixed collaboration structures with pre-designed agent roles, communication patterns, or workflow components \citep{li2023camel,wu2023autogen,hong2024metagpt,dat2026prism}. Yet coordination requirements vary across tasks: simple queries may need only one agent, independent subtasks can benefit from parallel execution, and dependent stages require reliable access to upstream results. Applying a fixed collaboration structure across these settings can introduce unnecessary communication, latency, and computation \citep{wang2026always,li2026sponge,yin2026policy}. We argue that scalable MAS require a stronger capability, which we refer to as \textit{Free Orchestration}: the ability to decide when collaboration is useful and construct task-specific workflows during problem solving.

Automated workflow design has begun to address this rigidity. Existing approaches search for effective workflows using execution feedback \citep{zhang2025aflow}, or generate and adapt query-specific agent systems during inference \citep{ye2025masgpt,gao2025flowreasoner,tieu2026inferencetime,wang2025dyflow,ke2025maszero,ke2026masorchestra}. However, these advances retain constraints on workflow construction and adaptation. DyFlow revises plans using execution feedback but constructs them from a finite set of operator templates \citep{wang2025dyflow}. MAS-Orchestra formulates orchestration through function calls, yet generates the complete coordination structure in a single decision without observing intermediate execution results \citep{ke2026masorchestra}. These limitations motivate combining free-form workflow synthesis with repeated orchestration decisions informed by execution feedback. We therefore investigate executable workflow synthesis as a native action of the task-solving agent, alongside ordinary tool use. The same agent decides (i) when to collaborate, (ii) constructs task-specific programs, and (iii) uses their returned outputs and traces to determine how to proceed.

To this end, we introduce \textbf{DynaFlow} (Figure~\ref{fig:overview}), a framework that integrates free-form executable workflow synthesis into the task-solving agent's action space. We refer to this agent as the \emph{main agent}, which coordinates \emph{sub-agents} responsible for executing delegated subtasks. The Main Agent can answer directly, use ordinary tools, or generate a task-specific workflow program. Through a minimal execution interface, it defines sub-agent responsibilities, sequential or parallel dependencies, and how intermediate outputs are passed, aggregated, and verified directly in code. A budget-constrained runtime executes the program and returns its outputs and execution trace. The Main Agent then uses this feedback to guide subsequent actions, including synthesizing follow-up workflows that adapt collaboration to the observed results.

We decouple orchestration learning from sub-agent training by updating only the Main Agent on distilled workflow demonstrations while keeping all sub-agents fixed. This focuses learning on coordinating existing capabilities and allows pretrained models to be reused as executors without retraining. The training process begins with distilled workflow demonstrations. Since answer correctness alone does not guarantee reliable coordination, we then refine the policy on a curated subset that excludes demonstrations flagged for unsupported premises, biased verification, or incomplete information transfer. This training process makes orchestration quality an explicit criterion for selecting demonstrations, guiding the Main Agent toward more reliable task coordination and information use.

Across six benchmarks spanning mathematical reasoning, scientific question answering, and multi-hop retrieval, DynaFlow achieves 88.08\% overall accuracy in Forced mode, exceeding the strongest evaluated baseline using the same backbone by 10.90 percentage points. In Dynamic mode, it maintains accuracy close to Forced execution while answering 47.1\% of questions without invoking a workflow. Further analysis finds that training leads to more explicit use, verification, and refinement of upstream outputs. Together, these results support learning both how to organize collaboration and when to invoke it, and identify cross-stage information use as a key dimension of effective orchestration.

\section{Related Work}
\label{sec:related-work}

\noindent\textbf{Agentic Workflows and Orchestration.}
A workflow specifies model and tool calls through instructions, control flow, and information dependencies. Control flow determines execution order and branching, while data flow specifies which intermediate outputs later calls receive. Orchestration organizes these calls to solve a task. AutoGen provides programmable agent interactions, while Magentic-One and Apodex~1.1 support runtime coordination and replanning \citep{wu2023autogen,fourney2024magenticone,apodex2026apodex11}. Graph- and code-based representations support automated workflow design \citep{zhuge2024gptswarm,zhang2025aflow}, and executable code can express tool-use actions and orchestration logic \citep{wang2024codeact,oconnor2026spe}. These advances provide expressive mechanisms for collaboration, while verification-centered systems emphasize checking intermediate decisions and their supporting evidence \citep{apodex2026apodex10,miromind2026mirothinker17}. We therefore focus on learning when to invoke collaboration and how to construct task-specific workflows that preserve, check, and reuse intermediate results. We further examine how training changes workflow patterns, focusing on stage dependencies and the use, verification, and refinement of intermediate results. Comparing these changes alongside design quality and task accuracy helps characterize the properties of effective orchestration and their relationship to improved performance (Figures~\ref{fig:structure} and~\ref{fig:quality}; \S\ref{sec:workflow-analysis}).

\noindent\textbf{Learning Task-Specific Orchestration.}
Task-specific orchestration raises two related questions: how to learn workflow generation and how to adapt designs using execution feedback, a mechanism also studied in agent--environment interaction and strategy-guided exploration \citep{miromind2025mirothinker,li2026apex}. MAS-GPT learns query-specific system generation through supervised training, while FlowReasoner and Mas-Orchestra use reinforcement learning \citep{ye2025masgpt,gao2025flowreasoner,ke2026masorchestra}. MAS-Zero revises system designs using intermediate results, and DyFlow plans successive subgraphs using feedback \citep{ke2025maszero,wang2025dyflow}. FlowSteer learns graph-editing policies from canvas feedback \citep{zhang2026flowsteer}. MetaAgent-X further jointly trains designers and executors \citep{zhang2026metaagentx}. DynaFlow integrates free-form executable workflow actions into the task-solving policy, coupling collaboration decisions with workflow construction and subsequent decisions based on execution feedback. Its learning framework explicitly selects demonstrations for orchestration quality and trains only the Main Agent, emphasizing reliable information transfer and independent verification with fixed sub-agents.

\section{DynaFlow}
\label{sec:method}

DynaFlow integrates free-form executable workflow synthesis into the Main Agent's action space. The Main Agent selects whether and how to collaborate, while workflow outputs and execution traces inform subsequent decisions (Figure~\ref{fig:overview}). We formalize this decision process before describing orchestration learning with fixed sub-agents.

\subsection{Free-Form Workflow Orchestration}
\label{sec:formulation}
\label{sec:runtime}
\label{sec:closed-loop}

Given a question $q$ and task instruction $I$, the Main Agent follows a policy $\pi_\theta$ over three action forms:
\begin{equation}
\mathcal{A}=\mathcal{A}_{\mathrm{tool}}\cup\{A_{\mathrm{ans}},(P,W(P))\},
\label{eq:action-space}
\end{equation}
where $\mathcal{A}_{\mathrm{tool}}$ is the set of ordinary tool-call actions, $A_{\mathrm{ans}}$ denotes a final-answer output action, and $(P,W(P))$ denotes a single workflow action comprising both the JavaScript program $P$ synthesized by the Main Agent and its invocation through the workflow tool $W$. Answer content and arguments for ordinary tools are generated with the corresponding actions and left implicit in the notation. At turn $k$, the Main Agent generates $a_k\sim\pi_\theta(\cdot\mid I,q,h_{k-1})$, where $h_{k-1}$ is the ordered history of previous actions and observations. The same policy therefore decides whether to answer, use a tool, or construct and invoke a workflow in light of the task and available feedback.

We evaluate this policy under two execution modes. In \textbf{Dynamic} mode, the first decision selects between a direct answer and a workflow call; in \textbf{Forced} mode, the first substantive action must invoke a workflow. Forced execution evaluates workflow construction without the initial routing decision. Both modes allow subsequent tool calls and follow-up workflows. Once a workflow is invoked, the Main Agent is restricted to coordination and summarization, with substantive problem solving delegated to sub-agents.

The program $P$ defines sub-agent roles, stages, and dependencies through \texttt{agent} calls, \texttt{parallel} execution, and optional \texttt{phase} labels. Independent calls can run concurrently, while dependent calls receive upstream outputs explicitly. Each sub-agent receives only the instructions and inputs specified by $P$. The workflow tool executes $P$ with fixed sub-agents under resource constraints, returning an observation $o_k=W(P)$ that combines the program output with an execution trace recording the status and output of each sub-agent call (Figure~\ref{fig:overview}, panel~3; Appendix~\ref{app:runtime}).

After receiving $o_k$, the Main Agent appends $(a_k,o_k)$ to the history to obtain $h_k$ and generates its next action conditioned on $I$, $q$, and $h_k$. When it judges the available information sufficient to answer $q$, it emits $A_{\mathrm{ans}}$ and terminates; otherwise, it invokes a tool or workflow to verify, refine, or extend the available results (Figure~\ref{fig:overview}, panel~4). Algorithm~\ref{alg:inference} summarizes this feedback loop over the action space in Eq.~\ref{eq:action-space}.

\subsection{Learning the Orchestration Policy}
\label{sec:training}

Building on the action formulation in \S\ref{sec:formulation}, we train the Main Agent policy $\pi_\theta$ to construct task-specific workflows and coordinate information flow across sub-agents. Training proceeds in two stages, updating only the Main Agent while keeping all sub-agents fixed (Figure~\ref{fig:overview}, panel~1).

\noindent\textbf{Stage 1: trajectory distillation.} A teacher model generates trajectories using the same workflow interface, tools, and fixed sub-agents. We retain trajectories with correct final answers that follow the delegation protocol in \S\ref{sec:formulation}, with substantive solving performed inside workflows. The supervised target $s=(r,a_1)$ is the Main Agent's first-turn output, where $r$ is its reasoning and $a_1=(P,W(P))$ is the workflow action defined in \S\ref{sec:formulation}. This yields a dataset $\mathcal{D}_1$ of examples $(I,q,s)$. Both training stages minimize
\begin{equation}
\mathcal{L}_{\mathrm{SFT}}(\theta;\mathcal{D})
= -\sum_{(I,q,s)\in\mathcal{D}}
\log\pi_\theta(r,a_1\mid I,q),
\label{eq:sft}
\end{equation}
with $\mathcal{D}=\mathcal{D}_1$ in Stage~1 and the curated subset $\mathcal{D}_2$ in Stage~2. The loss supervises both the reasoning tokens in $r$ and the action tokens in $a_1$; instructions, questions, and tool definitions are masked. Supervision thus targets first-turn task analysis and workflow synthesis.

\begin{algorithm}[t]
\caption{DynaFlow}
\label{alg:inference}
\begin{algorithmic}[1]
\Require $q$, $I$, $\pi_\theta$
\State $h_0\gets()$, $k\gets1$
\State $a_1\sim\pi_\theta(\cdot\mid I,q,h_0)$
\While{$a_k\ne A_{\mathrm{ans}}$}
  \If{$a_k=(P,W(P))$}
    \State $o_k\gets W(P)$
  \ElsIf{$a_k\in\mathcal{A}_{\mathrm{tool}}$}
    \State $o_k\gets\operatorname{Tool}(a_k)$
  \EndIf
  \State Append $(a_k,o_k)$ to $h_{k-1}$ to obtain $h_k$
  \State $a_{k+1}\sim\pi_\theta(\cdot\mid I,q,h_k)$
  \State $k\gets k+1$
\EndWhile
\State \Return $a_k$
\end{algorithmic}
\end{algorithm}

\begin{table*}[t]
\centering
\fontsize{9.3}{11.4}\selectfont
\setlength{\tabcolsep}{4pt}
\renewcommand{\arraystretch}{1.13}
\begin{tabularx}{0.9\textwidth}{l*{7}{>{\centering\arraybackslash}X}}
\arrayrulecolor{black}
\toprule
\rowcolor{DynaTableHeader}
\textbf{Method} &
\textbf{AIME24} &
\textbf{AIME25} &
\textbf{AIME26} &
\textbf{PolyMATH} &
\textbf{GPQA-D} &
\textbf{HotpotQA} &
\textbf{Overall} \\
\midrule
\multicolumn{8}{@{}l}{\textit{Single-agent baselines}} \\
CoT
& \underline{96.67} & 93.33 & 86.67 & 83.00 & 85.35 & 46.50 & 74.27 \\
CoT-SC
& 90.00 & 93.33 & \underline{96.67} & 84.50 & 85.86 & 50.50 & 76.16 \\
Self-Refine
& \underline{96.67} & 93.33 & 93.33 & 87.50 & \textbf{88.38} & 48.00 & 77.18 \\
\hline
\multicolumn{8}{@{}l}{\textit{Automated multi-agent baseline}} \\
AFlow
& 93.33 & \underline{96.67} & \textbf{100.00} & 86.00
& \underline{87.37} & 47.00 & 76.45 \\
\hline
\multicolumn{8}{@{}l}{\textit{Untrained Main Agent}} \\
Base, Direct
& 93.33 & 93.33 & 93.33 & 74.00 & 83.33 & 47.50 & 71.51 \\
Base + Web search
& 83.33 & 73.33 & 73.33 & 66.00 & 79.80 & 77.50 & 74.71 \\
Base, Dynamic
& 63.33 & 63.33 & 73.33 & 64.00 & 77.78 & 53.00 & 65.12 \\
Base, Forced
& 76.67 & 70.00 & 70.00 & 70.00 & 77.78 & 81.50 & 75.87 \\
\midrule
\multicolumn{8}{@{}l}{\textit{Trained Main Agent (ours)}} \\
\rowcolor{DynaTableHighlight}
\textbf{DynaFlow \textit{(Direct)}}
& 93.33 & 90.00 & 90.00
& 87.00 & 86.87 & 51.50 & 77.18 \\
\rowcolor{DynaTableHighlight}
\textbf{DynaFlow \textit{(Dynamic)}}
& 93.33 & \textbf{100.00} & \underline{96.67}
& \textbf{91.50} & 83.84 & \textbf{84.50} & \underline{87.94} \\
\rowcolor{DynaTableHighlight}
\textbf{DynaFlow \textit{(Forced)}}
& \textbf{100.00} & \textbf{100.00} & \textbf{100.00}
& \underline{90.50} & 84.85 & \underline{83.50} & \textbf{88.08} \\
\bottomrule[0.9pt]
\end{tabularx}
\caption{\textbf{Main results.} Accuracy (\%) on six benchmarks.
All systems use Qwen3.5-27B, and all workflow settings share the same
frozen sub-agents and runtime. Overall is micro-averaged over all
688 questions. Runs that end without a valid final answer count as
incorrect. Bold and underline mark the best and second-best result
in each column. 
with the single-agent baselines (\S\ref{sec:main-results}).}
\vspace{-5mm}
\label{tab:main}
\end{table*}

\noindent\textbf{Stage 2: high-quality data selection.} Stage~1 demonstrations meet the final-answer correctness and execution-protocol criteria while varying in orchestration quality. Evaluation of the Stage~1 model reveals that failures in information transfer and incorrect Main Agent judgments can steer workflows toward wrong answers. These observations motivate selecting high-quality demonstrations that support reliable information transfer and independent problem solving within workflows, with the aim of improving the Main Agent's orchestration capability.

Given each question $q$ and program $P$, an LLM judge assesses orchestration quality by checking for three types of defects: (i) \emph{premise injection}, where the Main Agent presents answer-determining claims beyond the task givens as established premises; (ii) \emph{verification defects}, where review instructions or output constraints steer a verifier toward an expected answer or restrict independent checking and revision; and (iii) \emph{information-flow defects}, where missing context, incorrect references, lossy transfer, or hard-coded substitutes for upstream outputs prevent required information from reaching downstream agents. Verification is assessed only when a reviewing step is present. Each reported defect must be supported by matching excerpts from the program or its launch arguments. We retain complete first-turn examples accepted after this evidence check. The judging rubric and agreement with manual annotation are described in Appendix~\ref{app:judge}. Denoting this selection procedure by $\mathcal{J}$, we obtain
\begin{equation}
\mathcal{D}_2=\mathcal{J}(\mathcal{D}_1)\subseteq\mathcal{D}_1.
\label{eq:curated}
\end{equation}
Stage~2 continues from the Stage~1 checkpoint on $\mathcal{D}_2$ with the same objective and a smaller learning rate. An equal-size random-subset continuation control is required to distinguish curation effects from additional training (\S\ref{sec:settings}). Training data and hyperparameters are detailed in \S\ref{sec:setup} and Appendix~\ref{app:training}.

\begin{figure}[t]
  \centering
  \includegraphics[width=\linewidth]{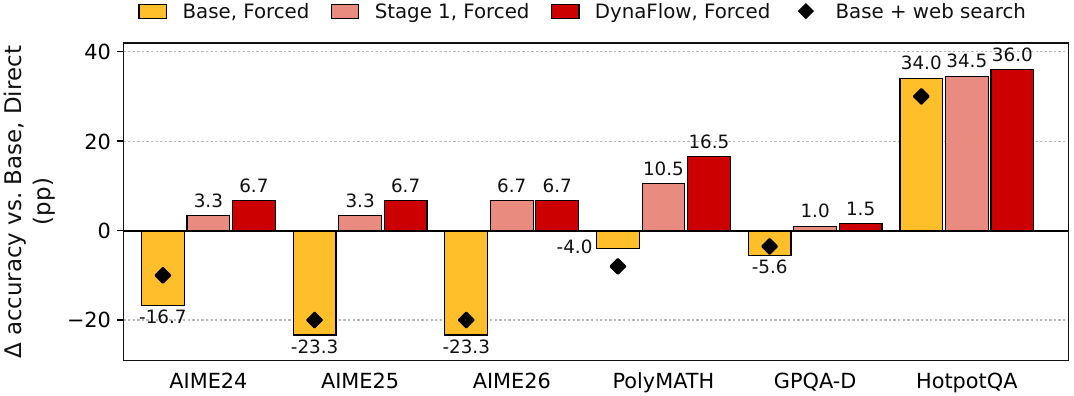}
  \caption{\textbf{Workflows help only after training.} Accuracy change in Forced mode relative to Base, Direct. Without training, workflows lower accuracy on every benchmark except HotpotQA; after training, Forced mode matches or exceeds Base, Direct on all six.}
  \label{fig:forced-vs-direct}
  \vspace{-5mm}
\end{figure}

\section{Experimental Setup}
\label{sec:setup}

Our experiments examine whether training the Main Agent improves workflow-based problem solving, how high-quality data selection affects orchestration, and whether adaptive workflow invocation preserves performance while reducing workflow use. Throughout the DynaFlow comparisons, sub-agents remain fixed, allowing us to study changes in the Main Agent under a shared workflow interface and tool configuration.

\noindent\textbf{Benchmarks and models.}
We evaluate on 688 questions across six benchmark sets: AIME 2024, 2025, and 2026, PolyMATH \citep{wang2025polymath} (200), GPQA-Diamond \citep{rein2023gpqa} (198), and HotpotQA \citep{yang2018hotpotqa} (200). The PolyMATH and HotpotQA subsets are sampled with random seed 42. These benchmarks cover competition mathematics, multilingual mathematical reasoning, scientific reasoning, and multi-hop question answering. HotpotQA uses web search for retrieval. The Main Agent uses Qwen3.5-27B \citep{qwen2026qwen35} as its backbone. Across all DynaFlow evaluation settings, workflow sub-agents uniformly use the Base Qwen3.5-27B checkpoint without orchestration training and remain fixed. Both the Main Agent and sub-agents operate with thinking enabled; only the Main Agent receives orchestration training.

\noindent\textbf{Training and evaluation settings.}
\label{sec:settings}
We compare three Main Agent checkpoints: \textbf{Base}, the backbone without orchestration training; \textbf{Stage~1}, trained by trajectory distillation; and \textbf{DynaFlow}, obtained after Stage~2 training on selected high-quality demonstrations. We use GLM-5.3 Max as the teacher model for trajectory distillation. Stage~1 uses QLoRA \citep{dettmers2023qlora} to fine-tune the Main Agent on 2{,}257 teacher-distilled first-turn examples; Stage~2 continues on 712 selected examples under Eq.~\eqref{eq:sft}.

The evaluation includes three execution modes: \textbf{Direct} answers without tools or workflows; \textbf{Dynamic} chooses between direct answering and workflow execution; and \textbf{Forced} requires workflow execution before any substantive problem solving. In both workflow-based paths, the Main Agent is restricted to designing and coordinating workflows, assessing returned results, and synthesizing the final answer; substantive solving and verification are delegated to sub-agents. Incomplete results trigger workflow revision or further calls (\S\ref{sec:formulation}).

\noindent\textbf{Baselines and metrics.}
\label{sec:baselines}
We compare against CoT, CoT-SC \citep{wang2023selfconsistency}, Self-Refine \citep{madaan2023selfrefine}, and AFlow \citep{zhang2025aflow}, all using Qwen3.5-27B. CoT-SC aggregates five sampled answers, while Self-Refine iteratively critiques and revises an answer. For AFlow, we transfer the released MATH workflow to AIME and PolyMATH, and the released HotpotQA workflow to HotpotQA and GPQA-Diamond, without searching for new workflows on our evaluation tasks. We also include \textbf{Base + Web search} to contextualize gains on retrieval-dependent questions. The tool-free baselines do not have the same retrieval access as the workflow settings; we therefore interpret their HotpotQA results separately from comparisons with search-enabled systems.

We report final-answer accuracy, counting runs without a valid final answer as incorrect. To examine orchestration behavior, we analyze Forced-mode workflows using structural indicators and four design criteria: task consistency, decomposition appropriateness, logical continuity, and information-flow correctness. Design scores range from 0 (major defect) through 1 (minor weakness) to 2 (no defect).

\begin{figure}[t]
  \centering
  \includegraphics[width=\linewidth]{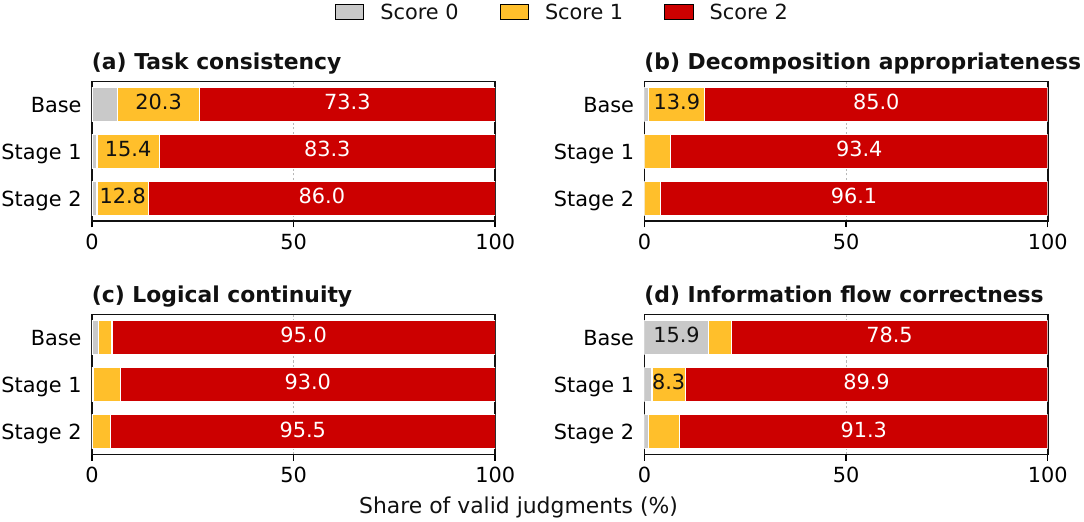}
  \caption{\textbf{Workflow design quality.} Judge scores for the first workflow generated for each question in Forced mode across all benchmarks: 0 = major defect, 1 = minor weakness, and 2 = no defect.}
  \label{fig:quality}
  \vspace{-5mm}
\end{figure}

\section{Results and Analysis}
\label{sec:results}

\subsection{Overall Performance}
\label{sec:main-results}

DynaFlow achieves the highest overall accuracy, reaching 88.08\% in Forced mode and 87.94\% in Dynamic mode (Table~\ref{tab:main}). Both exceed the strongest baseline by over 10 percentage points and outperform their Base counterparts, with only the Main Agent trained and all sub-agents fixed.

\noindent\textbf{Mathematical reasoning.}
DynaFlow achieves the strongest results across competition and multilingual mathematics. Forced execution solves all 90 AIME questions, achieving 100\% across the 2024, 2025, and 2026 sets, while both modes exceed the strongest baseline on PolyMATH by 3--4 percentage points. We analyze the generated workflows and find that they organize mathematical solving around complementary derivation, computation, and verification stages. For example, PolyMATH workflows pair symbolic derivations with small-instance numerical checks to identify errors in intermediate expressions. This integration offers a concrete explanation for the strong mathematical performance: intermediate results are exposed to checks that can guide subsequent correction.

\noindent\textbf{Scientific knowledge and reasoning.}
DynaFlow improves performance on GPQA-Diamond, which covers graduate-level questions in biology, chemistry, and physics. Forced execution reaches 84.85\%, improving over Base, Forced by 7.07 points and exceeding Base, Direct; Dynamic execution also improves over its Base counterpart. With the sub-agents fixed, we obtain stronger scientific question answering through Main Agent training, although Self-Refine and AFlow remain ahead on this benchmark. We examine failed cases and find that incorrect intermediate conclusions from sub-agents can misguide subsequent verification, while the Main Agent can fail to prioritize correct evidence when synthesizing the final answer. These failures highlight the importance of independent verification within workflows and reliable final-answer selection by the Main Agent.

\noindent\textbf{Multi-hop question answering with web search.}
DynaFlow achieves the best HotpotQA performance, reaching 84.50\% in Dynamic mode and 83.50\% in Forced mode. Both exceed Base + Web search by 6--7 points, and Forced execution also improves over Base, Forced under the same workflow requirement. We analyze the retrieval workflows and find that they organize search around intermediate entities and verify the connections between successive hops. For example, a workflow first identifies the city associated with a venue, then retrieves the city's nickname and checks that both hops refer to the same entity. Such evidence-chain construction and verification provide a plausible account of the gains over a search-enabled baseline and illustrate how DynaFlow organizes retrieval into a coherent solving process.

\subsection{Effect of Training and Execution Mode}
\label{sec:stage-results}

\begin{table}[t]
\centering
\small
\setlength{\tabcolsep}{4.5pt}
\begin{tabular}{lccc}
\arrayrulecolor{black}
\toprule
\rowcolor{DynaTableHeader}
\textbf{Main Agent}
& \textbf{Direct}
& \textbf{Dynamic}
& \textbf{Forced} \\
\midrule
Base
& 71.51 & 65.12 & 75.87 \\
Stage~1
& 76.74 & 84.45 & 85.47 \\
\rowcolor{DynaTableHighlight}
\textbf{DynaFlow (Stage~2)}
& 77.18 & \underline{87.94} & \textbf{88.08} \\
\bottomrule
\end{tabular}
\caption{%
\textbf{Training stage $\times$ execution mode.}
Overall accuracy (\%).
Only the Main Agent differs between rows.
Per-benchmark results are in Table~\ref{tab:stage-full}.}
\label{tab:stage}
\vspace{-5mm}
\end{table}

\begin{figure*}[t]
  \centering
  \includegraphics[width=0.85\textwidth]{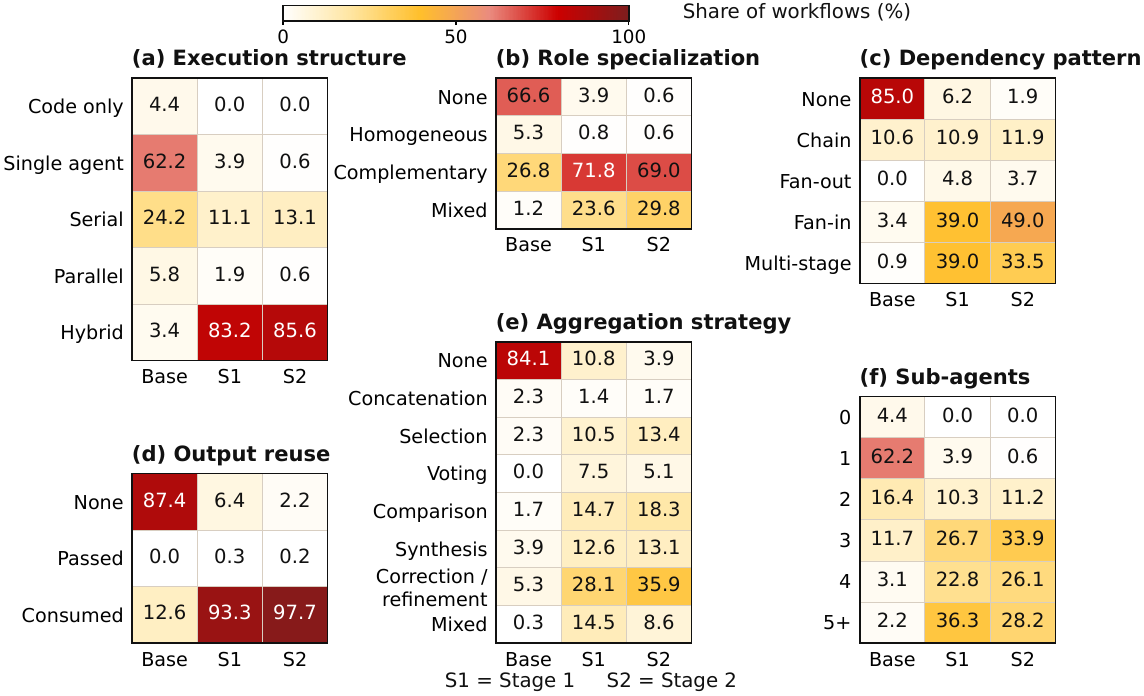}
  \caption{\textbf{\textbf{Workflow structure and information use across stages.}} Distributions (\%) of six properties of workflow programs $P$ generated under Forced execution: (a) execution structure, (b) role specialization, (c) dependency pattern, (d) output reuse, (e) aggregation strategy, and (f) sub-agent count. Definitions are provided in Table~\ref{tab:workflow-structure-definitions}.}
  \label{fig:structure}
  \vspace{-4mm}
\end{figure*}

We examine how training affects workflow-based solving under matched execution settings (Table~\ref{tab:stage}; Figure~\ref{fig:forced-vs-direct}). Base, Forced underperforms Base, Direct on every benchmark except HotpotQA, where workflows enable retrieval. After Stage~1, Forced execution reaches or exceeds the Base, Direct reference on all six benchmarks and surpasses Direct within the same checkpoint by 8.73 points overall, including retrieval gains on HotpotQA. We thus find that orchestration training turns workflow access into effective collaboration, yielding gains beyond the trained Main Agent's standalone performance. This pattern is consistent with prior findings that collaboration benefits depend on task characteristics and coordination structure \citep{kim2026capable}.

Stage~2 further improves both Forced and Dynamic performance. Our workflow analysis finds higher design quality and more frequent use and refinement of upstream results with fewer sub-agents on average than Stage~1 (Figures~\ref{fig:quality} and~\ref{fig:structure}; \S\ref{sec:workflow-analysis}), linking the performance gains to more focused coordination.

\subsection{Workflow Quality and Task-Specific Organization}
\label{sec:workflow-analysis}

\noindent\textbf{Connected stages and improved information use.}
Figure~\ref{fig:structure}(a,c) shows that Base typically delegates to a single sub-agent or constructs calls without dependencies, whereas trained programs usually combine parallel and sequential stages. Hybrid workflows increase from 3.4\% for Base to 83.2\% after Stage~1 and 85.6\% after Stage~2. The share of programs that pass upstream outputs and explicitly instruct downstream agents to use them rises from 12.6\% for Base to 93.3\% after Stage~1 and 97.7\% after Stage~2 (Figure~\ref{fig:structure}(d)). Training thus improves information use in workflow design: the Main Agent increasingly specifies how intermediate results should inform subsequent solving and verification, enabling systematic reuse across stages.

\begin{figure*}[t]
  \centering
  \includegraphics[width=0.75\textwidth]{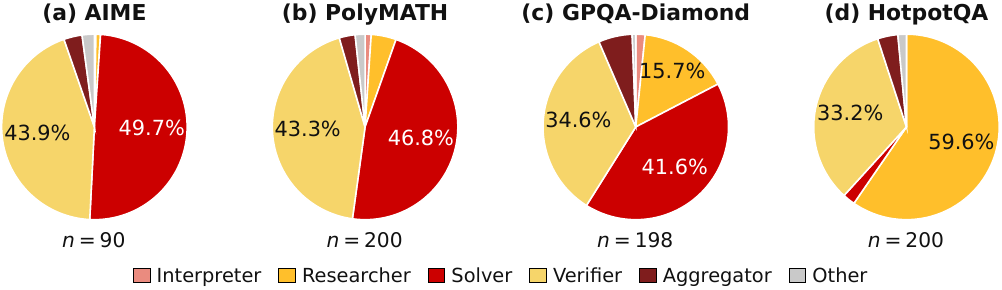}
 \caption{\textbf{Task-specific role allocation.} Percentages of executed sub-agent nodes by role and benchmark family in DynaFlow's Forced mode. The three AIME years are pooled; $n$ denotes question counts. Classification details are in Appendix~\ref{app:roles}.}
  \label{fig:roles}
  \vspace{-4mm}
\end{figure*}
\noindent\textbf{Higher-quality information flow and verification.}
Our analysis of Stage~1 errors identifies three recurring patterns: unsupported premises introduced by the Main Agent, verification guided toward an expected conclusion, and incomplete or incorrect transfer of intermediate information. These observations motivate selecting Stage~2 demonstrations that avoid premise injection, verification defects, and information-flow defects (\S\ref{sec:training}). After training on the selected data, we find higher proportions of defect-free workflows across all four design criteria. Major information-flow defects decline from 1.9\% in Stage~1 to 0.9\% in Stage~2 (Figure~\ref{fig:quality}), while workflows that specify correction or refinement of intermediate results increase from 28.1\% to 35.9\% (Figure~\ref{fig:structure}(e)). These results indicate better workflow design for transferring, integrating, and checking intermediate results through explicit dependencies and structured information exchange, with fewer sub-agents on average (4.16 versus 4.35 in Stage~1). We interpret Base's relatively high logical-continuity scores in light of its simpler, often single-agent workflows, which involve fewer dependencies between stages.

\noindent\textbf{Task-specific roles and instructions.}
Figure~\ref{fig:roles} shows distinct role configurations across tasks. Solvers and verifiers account for over 90\% of nodes on AIME and PolyMATH, while researchers account for 59.6\% on HotpotQA. GPQA-Diamond combines a substantial solver component with more research than the mathematics benchmarks. Verification remains common across tasks, accounting for 33.2--43.9\% of nodes. This task-specific organization extends to the instructions themselves: the Main Agent generates a prompt for each sub-agent that specifies its objective, relevant inputs, and use of upstream results for the current question (\S\ref{sec:formulation}). DynaFlow thus adapts both the division of labor and the instructions that guide execution, connecting task decomposition with information use. Relative to the transferred AFlow workflows used in our evaluation (\S\ref{sec:baselines}), this flexibility supports question-specific mathematical derivation, evidence gathering, and independent checking (\S\ref{sec:main-results}).

\subsection{Adaptive Workflow Invocation}
\label{sec:adaptive-invocation}

The Dynamic--Forced accuracy gap narrows from 10.75 percentage points for Base to 1.02 after Stage~1 and 0.14 for DynaFlow (Table~\ref{tab:stage}). With sub-agents fixed, this trend highlights the contribution of Main Agent training to selective collaboration. In Dynamic mode, DynaFlow answers 47.1\% of questions directly while achieving 87.94\% accuracy, close to 88.08\% under Forced execution. Averaged over all questions, including those answered directly, it uses 2.44 sub-agents per question versus approximately 4.16 under Forced execution, a reduction of about 41\% (Figure~\ref{fig:teaser}). These results suggest that a unified Main Agent policy can adapt its solution strategy to task requirements while maintaining performance with fewer sub-agents. We regard the choice between direct answering and workflow execution as an integral part of orchestration: additional agents do not necessarily improve performance, and collaboration should be invoked according to the needs of each question.

\section{Discussion}
\label{sec:discussion}

We view reliable information use alongside task decomposition as central to orchestration; gains with fixed sub-agents support learning to coordinate existing capabilities (\S\ref{sec:stage-results}). Stage~2 gains accompany better information transfer and more frequent refinement of intermediate results, consistent with screening for unsupported premises, constrained verification, and information-flow defects (\S\ref{sec:workflow-analysis}). Our analysis of GPQA-Diamond failures illustrates how correct intermediate results can be lost or overridden during verification and final synthesis. DynaFlow workflows incorporate complementary evidence gathering, independent checks, and refinement of intermediate results. Future work could train sub-agents for reliable information transfer and independent verification, and extend first-turn supervision to multi-turn trajectories containing disagreements and erroneous results. Such supervision could strengthen the Main Agent's ability to independently assess conflicting outputs, identify errors, and select actions for error recovery, workflow revision, and final-answer selection.

\section{Conclusion}
\label{sec:conclusion}
DynaFlow combines free-form executable workflow actions with two-stage Main Agent training through trajectory distillation and demonstration selection based on orchestration quality, keeping sub-agents fixed. Across six benchmarks, it achieves the highest overall accuracy among evaluated methods; Dynamic execution answers nearly half the questions directly with accuracy close to Forced execution. Workflow analysis characterizes how training changes structure and the use of intermediate results.





\bibliography{custom}
\clearpage
\appendix
\flushbottom

\captionsetup{font=small,skip=6pt}
\setlength{\textfloatsep}{14pt plus 3pt minus 3pt}
\setlength{\dbltextfloatsep}{14pt plus 3pt minus 3pt}
\setlength{\floatsep}{12pt plus 2pt minus 2pt}
\setlength{\dblfloatsep}{12pt plus 2pt minus 2pt}
\renewcommand{\topfraction}{0.9}
\renewcommand{\dbltopfraction}{0.9}
\renewcommand{\textfraction}{0.07}
\renewcommand{\floatpagefraction}{0.7}
\renewcommand{\dblfloatpagefraction}{0.7}
\makeatletter
\setlength{\@fptop}{0pt}
\setlength{\@dblfptop}{0pt}
\setlength{\@fpsep}{12pt plus 2pt minus 2pt}
\setlength{\@dblfpsep}{12pt plus 2pt minus 2pt}
\newcommand{\DynaAppendixClearPage}{%
  \clearpage
  \global\@colroom\@colht
  \global\vsize\@colht
}
\makeatother
\providecolor{DynaTableHeader}{HTML}{FFF1CC}
\providecolor{DynaTableHighlight}{HTML}{FBECEC}

\newcommand{\DynaAppendixFigure}[4][0.72\textheight]{%
  \begin{figure*}[!tp]
    \centering
    \includegraphics[width=\textwidth,height=#1,keepaspectratio]{#2}
    \caption{#3}
    \label{#4}
  \end{figure*}
}

\definecolor{DynaPromptBlue}{HTML}{285E8E}
\definecolor{DynaPromptTeal}{HTML}{187A75}
\definecolor{DynaPromptAmber}{HTML}{A86719}
\definecolor{DynaPromptPurple}{HTML}{68528C}
\definecolor{DynaPromptSlate}{HTML}{4E5C6D}
\newcommand{\DynaAppendixPromptCard}[3]{%
  \par\addvspace{7pt}\begingroup
  \setlength{\fboxsep}{5pt}%
  \setlength{\fboxrule}{0.5pt}%
  \noindent\fcolorbox{#1}{#1!4}{%
    \begin{minipage}{\dimexpr\linewidth-2\fboxsep-2\fboxrule\relax}
      \fontsize{10}{12}\selectfont\raggedright
      \setlength{\parindent}{0pt}\setlength{\parskip}{3pt}
      \setlength{\emergencystretch}{1em}%
      \noindent\colorbox{#1}{%
        \parbox{\dimexpr\linewidth-2\fboxsep\relax}{%
          \color{white}\bfseries\strut #2}}\par\nobreak
      #3\par
    \end{minipage}}\par
  \endgroup}

\section{Experimental Setup Details}
\label{app:setup}

\paragraph{Benchmarks.}
The evaluation contains 688 questions: 30 from each of AIME 2024, 2025, and 2026; 200 from PolyMATH; 198 from GPQA-Diamond; and 200 from HotpotQA. PolyMATH and HotpotQA are sampled with seed 42. AIME requires integer answers, PolyMATH includes multilingual mathematical questions with free-form answers, GPQA-Diamond uses multiple-choice scientific questions, and HotpotQA requires connecting evidence across retrieval steps.

\paragraph{Models and baselines.}
The Agent uses Qwen3.5-27B, and all workflow sub-agents use its Base checkpoint without orchestration training. Both operate with thinking enabled; only the Agent receives orchestration training. CoT, five-sample CoT-SC, Self-Refine, and AFlow use the same backbone. AFlow uses the released MATH workflow on AIME and PolyMATH and the released HotpotQA workflow on HotpotQA and GPQA-Diamond. These workflows are transferred without an additional workflow search on the evaluation tasks. Base + Web search provides a comparison with direct retrieval access.

\section{Workflow Runtime}
\label{app:runtime}

The Main Agent passes a JavaScript program to the Workflow tool. Programs define sub-agent instructions and pass intermediate outputs through ordinary code. The interface provides \texttt{agent} for sub-agent calls, \texttt{parallel} for concurrent calls, \texttt{pipeline} for dependent stages over multiple items, and \texttt{phase} for stage labels. A sub-agent receives the instruction and inputs specified by the program; the Main Agent's interaction history is not automatically inherited. Calls may request structured outputs through a JSON schema. Workflow completion returns the program result together with execution information, which the Main Agent can use to revise the workflow or produce a final answer. During evaluation, Direct answers without tools, Dynamic selects between direct answering and workflow execution, and Forced requires workflow invocation before substantive problem solving. The corresponding prompts are provided in Appendix~\ref{app:modes}.

\section{Training Details}
\label{app:training}

GLM-5.3 Max provides the teacher trajectories. The Stage~1 training set contains 2{,}257 first-turn demonstrations from HotpotQA and BrowseComp. Supervision covers the first-turn reasoning and action; instructions, questions, and tool definitions are masked from the loss. Only the Main Agent is fine-tuned, while all sub-agents remain fixed.

Stage~2 continues from the Stage~1 checkpoint on 712 complete workflow demonstrations selected from the Stage~1 training set: 449 from HotpotQA and 263 from BrowseComp. Both stages train for two epochs using the objective in Eq.~\eqref{eq:sft}, with a smaller learning rate in Stage~2 and no separate validation split. Table~\ref{tab:training} summarizes the training configuration, and Appendix~\ref{app:judge} describes data selection.

We use \texttt{ms-swift}~4.6.0.dev0 for training with 4-bit NF4 QLoRA, double quantization, BF16 computation, and gradient checkpointing. The Qwen3.5-27B backbone has approximately 27 billion parameters, with 116{,}727{,}808 trainable parameters in our configuration. LoRA targets eligible linear layers through \texttt{all-linear}, while the visual encoder and alignment module remain frozen. Each stage runs on two NVIDIA H100 80GB GPUs, with a single training process sharding the model across both devices. Based on full job wall-clock time, Stage~1 and Stage~2 consume about 52 and 17 GPU-hours, respectively, totaling about 69 GPU-hours. This budget covers fine-tuning only, excluding teacher trajectory generation, evaluation, and LLM-based judging.

\begin{table}[!htbp]
\centering
\small
\setlength{\tabcolsep}{4pt}
\renewcommand{\arraystretch}{1.12}
\begin{tabularx}{\linewidth}{@{}>{\raggedright\arraybackslash}Xcc@{}}
\toprule
\textbf{Parameter} & \textbf{Stage 1} & \textbf{Stage 2} \\
\midrule
Training examples & 2{,}257 & 712 \\
Learning rate & $5\times10^{-5}$ & $1\times10^{-5}$ \\
Epochs & 2 & 2 \\
\midrule
LoRA rank / $\alpha$ / dropout & \multicolumn{2}{c}{16 / 32 / 0} \\
Optimizer & \multicolumn{2}{c}{Fused AdamW} \\
Schedule / warmup ratio & \multicolumn{2}{c}{Cosine / 0.05} \\
Effective batch size & \multicolumn{2}{c}{4} \\
Max sequence length & \multicolumn{2}{c}{67{,}584} \\
Random seed & \multicolumn{2}{c}{42} \\
\bottomrule
\end{tabularx}
\caption{\textbf{Training configuration.} Only the Main Agent is updated. Shared values apply to both stages; maximum sequence length denotes the configured upper limit.}
\label{tab:training}
\end{table}

\section{Evaluation Protocol}
\label{app:evaluation}

\paragraph{Decoding configuration.}
Table~\ref{tab:runtime} summarizes the decoding configuration for DynaFlow evaluation. Thinking is enabled for both the Main Agent and sub-agents. The output-token limits apply to individual model responses, rather than the number of workflow stages or interaction turns.

\begin{table}[!htbp]
\centering
\small
\setlength{\tabcolsep}{4pt}
\renewcommand{\arraystretch}{1.12}
\begin{tabularx}{\linewidth}{@{}>{\raggedright\arraybackslash}Xr@{}}
\toprule
\textbf{Parameter} & \textbf{Value} \\
\midrule
\multicolumn{2}{@{}l}{\textit{Main Agent}} \\
Context length & 131{,}072 \\
Max output tokens & 81{,}920 \\
Temperature / top-$p$ / top-$k$ & 1.0 / 0.95 / 20 \\
Presence penalty / seed & 1.5 / 42 \\
\midrule
\multicolumn{2}{@{}l}{\textit{Sub-agents}} \\
Max output tokens & 65{,}536 \\
\bottomrule
\end{tabularx}
\caption{\textbf{Evaluation decoding parameters.}}
\label{tab:runtime}
\end{table}

\paragraph{Answer scoring.}
For DynaFlow, the prediction is extracted from the Main Agent's final \texttt{FINAL ANSWER} line; intermediate reasoning and sub-agent outputs are not substituted for an absent final answer. Runs without a valid final answer are counted as incorrect. AIME answers are matched as integers after normalizing leading zeros, and GPQA-Diamond answers are matched to the question's option labels. PolyMATH permits exact mathematical equivalence, while HotpotQA accepts answers that identify the reference entity or relation, including equivalent names and aliases. Approximate decimals are not accepted as exact answers when the task specifies no numerical precision.

\paragraph{Workflow quality and structure.}
\label{app:structure}
Workflow programs generated in Forced mode are evaluated by \texttt{deepseek-flash} \citep{deepseekai2026deepseekv41flash} with thinking enabled, high reasoning effort, and temperature 0, using the question and program without execution results. Task consistency, decomposition appropriateness, logical continuity, and information-flow correctness are scored as 0 (major defect), 1 (minor weakness), or 2 (no defect). Structurally inapplicable logical-continuity and information-flow ratings are mapped to 2 when no relevant dependencies are required; insufficient-evidence ratings are excluded. We characterize the workflow programs $P$ generated by the Main Agent along six structural dimensions (Table~\ref{tab:workflow-structure-definitions}) and compare the distributions of these properties across training stages in Figure~\ref{fig:structure}.

\begin{table*}[!tp]
\centering
\small
\setlength{\tabcolsep}{5pt}
\renewcommand{\arraystretch}{1.1}
\begin{tabularx}{\textwidth}{>{\raggedright\arraybackslash}p{0.17\textwidth}>{\raggedright\arraybackslash}X}
\toprule
\rowcolor{DynaTableHeader}
\textbf{Dimension} & \textbf{Categories and definitions} \\
\midrule
Execution structure &
\textbf{Code only}: no agent reasoning node, only code or tools;
\textbf{Single agent}: exactly one agent reasoning node;
\textbf{Serial}: multiple agents operate mainly in sequence, with later stages depending on earlier ones;
\textbf{Parallel}: agents operate mainly independently in parallel, without substantial sequential dependencies;
\textbf{Hybrid}: parallel branches and sequential agent dependencies are meaningfully combined. \\
\addlinespace[3pt]
Role specialization &
\textbf{None}: fewer than two agents;
\textbf{Homogeneous}: agents mainly perform the same role or type of attempt;
\textbf{Complementary}: agents differ in roles, perspectives, subtasks, or methods;
\textbf{Mixed}: homogeneous attempts and complementary roles coexist. \\
\addlinespace[3pt]
Dependency pattern &
\textbf{None}: no agent depends on another agent's output;
\textbf{Chain}: outputs mainly form a sequential chain;
\textbf{Fan-out}: one upstream result feeds multiple downstream agents;
\textbf{Fan-in}: one downstream agent combines multiple upstream results;
\textbf{Multi-stage}: multiple dependency forms or sequential stages are combined. \\
\addlinespace[3pt]
Output reuse &
\textbf{None}: no upstream output is passed downstream;
\textbf{Passed}: an upstream output is supplied, but its meaningful use is not explicitly required;
\textbf{Consumed}: the actual upstream output is supplied and the downstream instruction explicitly requires using, analyzing, comparing, refining, critiquing, or transforming it. \\
\addlinespace[3pt]
Aggregation strategy &
\textbf{None}: no result aggregation;
\textbf{Concatenation}: results are collected without substantive comparison or synthesis;
\textbf{Selection}: one result or candidate is selected;
\textbf{Voting}: results are combined by voting or majority agreement;
\textbf{Comparison}: outputs are explicitly compared;
\textbf{Synthesis}: outputs are integrated into a new answer or representation;
\textbf{Correction/refinement}: an upstream result is reviewed and may be revised;
\textbf{Mixed}: multiple substantive aggregation mechanisms are used. \\
\addlinespace[3pt]
Agent count &
Number of distinct agent reasoning nodes explicitly instantiated in the program, excluding deterministic code, formatting, and simple tool calls. Counts are displayed as 0, 1, 2, 3, 4, and 5+ in Figure~\ref{fig:structure}. \\
\bottomrule
\end{tabularx}
\caption{\textbf{Workflow structure definitions.} Six structural dimensions used to characterize the workflow programs $P$ generated by the Main Agent.}
\label{tab:workflow-structure-definitions}
\end{table*}

\paragraph{Role assignment.}
\label{app:roles}
Figure~\ref{fig:roles} summarizes the roles of sub-agents executed in each DynaFlow, Forced run, excluding the Main Agent. Based on commonly occurring roles in the generated workflows, we define five broad categories: Verifier, Aggregator, Interpreter, Researcher, and Solver. Categories are assigned using keywords in node labels, with phase names as a fallback. When multiple categories match, the order above determines priority; unmatched nodes are labeled Other. The Aggregator category covers synthesis, merging, and adjudication. Role proportions are computed over executed sub-agent nodes, with results pooled across the AIME benchmarks.

\section{Orchestration Quality Selection}
\label{app:judge}

The selection judge is \texttt{deepseek-flash}, with thinking enabled, high reasoning effort, and temperature 0. It receives the question, the workflow program generated in the first turn, and any launch arguments, without the Main Agent's reasoning, final answer, reference answer, or execution results. The rubric checks three defects: \emph{premise injection}, where unsupported answer-relevant claims are supplied as established facts; \emph{verification defects}, where instructions or output constraints favor an expected conclusion or prevent independent correction; and \emph{information-flow defects}, where missing context, incorrect references, or substituted values prevent required information from reaching downstream agents.

The judge cites excerpts from the workflow program or launch arguments to support each reported defect. We verify these excerpts against the supplied text as part of the quality screening. From the 2{,}257 workflow-containing demonstrations, we retain 712 complete first-turn examples for Stage~2.

\section{Evaluation Prompts and Workflow Interface}
\label{app:modes}
\label{app:prompts}

Figure~\ref{fig:appendix-prompts} presents concise versions of the Direct, Dynamic, and Forced evaluation prompts, the shared answer format, and the Workflow tool interface. These prompts specify the conditions for workflow invocation, task delegation, and use of returned results. The Main Agent determines sub-agent roles, instructions, and workflow structure for each question. Repeated formatting instructions are omitted for brevity.

\begin{figure*}[!tp]
\centering
\noindent\begin{minipage}[t]{0.49\linewidth}\vspace{0pt}
\DynaAppendixPromptCard{DynaPromptBlue}{Direct: task prompt}{%
You are a professional AI assistant. Solve the task using only the supplied information, your existing knowledge, and reasoning, without any task-solving tool calls.

This boundary applies from the beginning of the task and throughout execution, including search, web reading, and code execution. For attachments, use the file contents supplied with the task; report the limitation if required contents are unavailable.
}

\DynaAppendixPromptCard{DynaPromptTeal}{Dynamic: task prompt}{%
Before substantive problem solving, select one execution path:

\textbf{DIRECT:} Answer using the task information, existing knowledge, and reasoning, without task-solving tools.

\textbf{WORKFLOW:} Organize the solution through Workflow. Substantive research, calculations, code execution, and verification must occur within the Workflow.

Do not solve the task before selecting the path. If retrieval or another task-solving tool is needed, select WORKFLOW before the first such call. If DIRECT is initially selected but tools are later needed, switch to WORKFLOW first.
}
\end{minipage}\hfill
\begin{minipage}[t]{0.49\linewidth}\vspace{0pt}
\DynaAppendixPromptCard{DynaPromptAmber}{Forced: task prompt}{%
Invoke Workflow before substantive problem solving, even when the task could be answered from existing knowledge. Research, calculations, code execution, reasoning, and verification must occur within the Workflow.

The Main Agent may only plan, launch, monitor, revise, assess returned results, and summarize the final answer. This includes the first reasoning turn: understand the task and design its execution, without independently deriving the answer or supplying sub-agents with a pre-solved answer.

A completion status alone is insufficient when required subtasks have not returned results. Treat the run as incomplete and revise or rerun the Workflow.
}

\DynaAppendixPromptCard{DynaPromptSlate}{Shared task and answer format}{%
The task supplies a question and task identifier, with optional attachment names or paths.

Before the final answer, output a record:

\texttt{\{"task\_id": "<task\_id>",}\par
\texttt{\hspace*{1em}"Final answer": "<final\_answer>"\}}

If no task identifier is supplied, use an empty string. Finish the response with:

\texttt{FINAL ANSWER: [YOUR FINAL ANSWER]}

Follow the answer format requested by the task and do not include explanations after this final line.
}
\end{minipage}\par

\DynaAppendixPromptCard{DynaPromptPurple}{Workflow: delegation and tool interface}{%
\textbf{Task delegation.} Provide the original task, relevant context, attachment paths, and output requirements. Design the Workflow for the current task, using sub-agents, parallelism, verification, and additional stages as appropriate. Read and follow the tool description and input schema before invocation.

\textbf{Invocation.} Supply a JavaScript program through \texttt{script}, or a stored program through \texttt{scriptPath}, with optional \texttt{args}. Each script begins with a literal \texttt{meta} object containing \texttt{name} and \texttt{description}; optional \texttt{phases} describe its stages. The tool starts execution in the background and returns a task identifier; a completion notification provides the program's returned result.

\textbf{Execution primitives.} \texttt{agent(prompt, options)} launches a sub-agent and returns final text or a validated object when a JSON schema is supplied. \texttt{parallel(thunks)} runs calls concurrently and waits for their results. \texttt{pipeline(items, stages...)} processes each item through dependent stages without waiting for unrelated items. \texttt{phase(title)} labels a stage, and \texttt{args} exposes invocation inputs. Ordinary JavaScript passes and transforms intermediate outputs in an asynchronous context.

\textbf{Result handling.} Do not independently duplicate the Workflow's problem solving in the Main Agent. If results are incomplete, uncertain, or contradicted by evidence, revise or rerun the Workflow. Answer only when the returned results are sufficient, and base the final answer on them. Report a failure if Workflow is unavailable or cannot complete.
}
\caption{\textbf{Evaluation prompts and Workflow interface.} Abridged prompt excerpts for the three execution modes, the shared answer format, and the Workflow tool.}
\label{fig:appendix-prompts}
\end{figure*}

\section{Full Results}
\label{app:full-results}

Table~\ref{tab:stage-full} lists per-benchmark accuracy for each Main Agent and execution mode. Figure~\ref{fig:structure} provides the detailed structural distributions underlying the workflow analysis in \S\ref{sec:workflow-analysis}.

\begin{table*}[!tp]
\centering
\small
\setlength{\tabcolsep}{3pt}
\renewcommand{\arraystretch}{1.12}
\begin{tabularx}{\textwidth}{ll*{7}{>{\centering\arraybackslash}X}}
\arrayrulecolor{black}
\toprule
\rowcolor{DynaTableHeader}
\textbf{Main Agent}
& \textbf{Mode}
& \textbf{AIME24}
& \textbf{AIME25}
& \textbf{AIME26}
& \textbf{PolyMATH}
& \textbf{GPQA-D}
& \textbf{HotpotQA}
& \textbf{Overall} \\
\midrule
Base & Direct
& 93.33 & 93.33 & 93.33
& 74.00 & 83.33 & 47.50 & 71.51 \\
& Dynamic
& 63.33 & 63.33 & 73.33
& 64.00 & 77.78 & 53.00 & 65.12 \\
& Forced
& 76.67 & 70.00 & 70.00
& 70.00 & 77.78 & 81.50 & 75.87 \\
\midrule
Stage~1 & Direct
& 93.33 & 96.67 & 90.00
& 87.50 & 87.88 & 47.50 & 76.74 \\
& Dynamic
& 96.67 & 100.00 & 93.33
& 88.50 & 82.83 & 76.50 & 84.45 \\
& Forced
& 96.67 & 96.67 & 100.00
& 84.50 & 84.34 & 82.00 & 85.47 \\
\midrule
\rowcolor{DynaTableHighlight}
\textbf{DynaFlow} & Direct
& 93.33 & 90.00 & 90.00
& 87.00 & 86.87 & 51.50 & 77.18 \\
\rowcolor{DynaTableHighlight}
& Dynamic
& 93.33 & 100.00 & 96.67
& 91.50 & 83.84 & 84.50 & 87.94 \\
\rowcolor{DynaTableHighlight}
& Forced
& 100.00 & 100.00 & 100.00
& 90.50 & 84.85 & 83.50 & 88.08 \\
\bottomrule
\end{tabularx}
\caption{%
\textbf{Results by training stage and execution mode.}
Per-benchmark accuracy (\%) by Main Agent and execution mode.
DynaFlow is Stage~1 followed by curated training (Stage~2).
Shaded rows denote DynaFlow.}
\label{tab:stage-full}
\end{table*}

\DynaAppendixClearPage
\section{Case Studies of Workflow Changes}
\label{app:workflow-cases}

We compare selected Base and DynaFlow workflows under Forced execution with fixed sub-agents, complementing the analysis of information use and task-specific organization in \S\ref{sec:workflow-analysis}. Each diagram is reconstructed from the workflow program and execution record; captions specify the displayed invocation and distinguish executed calls from conditional branches.

\paragraph{Complementary computation and explicit verification.}
For the palindrome-counting problem from AIME 2026 (Figure~\ref{fig:case-aime260}), Base delegates to one solver, which returns 59 after an arithmetic error. DynaFlow assigns enumeration, dynamic programming, and inclusion--exclusion to parallel solvers, then passes their outputs to a verifier for comparison and independent recomputation. All three solvers and the verifier obtain 62. This case illustrates how complementary methods and explicit use of intermediate results support mathematical verification.

\DynaAppendixFigure
  {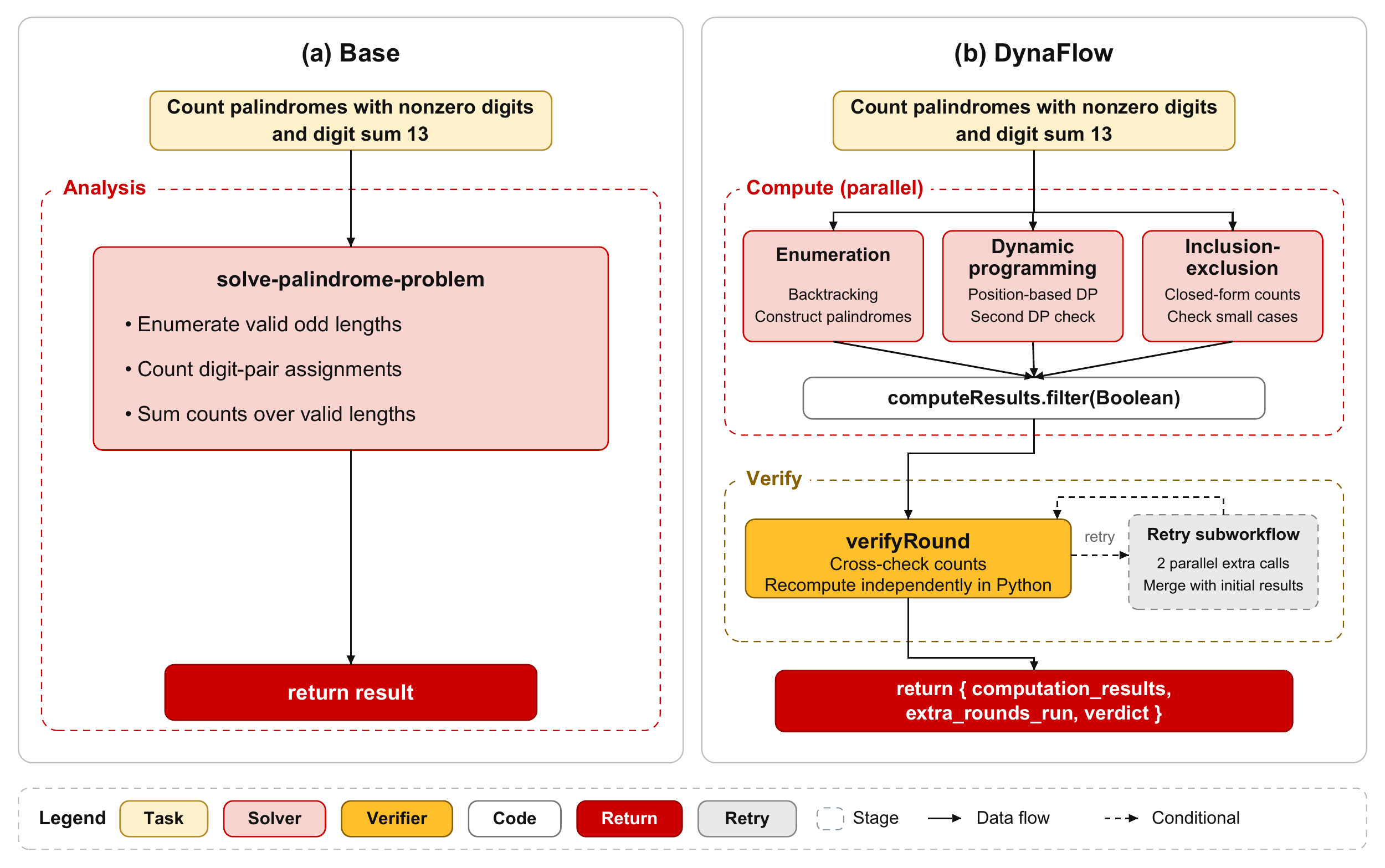}
  {\textbf{AIME 2026: palindrome counting.} Base uses one solver; DynaFlow connects three complementary computations to a common verifier. The dashed retry branch was not triggered. Repeated input and result packaging is omitted.}
  {fig:case-aime260}

\paragraph{Preserving research evidence for scientific adjudication.}
For the reaction-product question from GPQA-Diamond (Figure~\ref{fig:case-gpqa77}), Base runs two analyses sequentially without passing results between them. DynaFlow assigns four research calls to distinct chemical constraints and checks the candidate products for each reaction. Both checks challenge their initial candidates, and the decision agent receives the research evidence alongside the verification results. Preserving these complementary inputs enables final synthesis to account for both the proposed answers and the objections raised during verification.

\DynaAppendixFigure
  {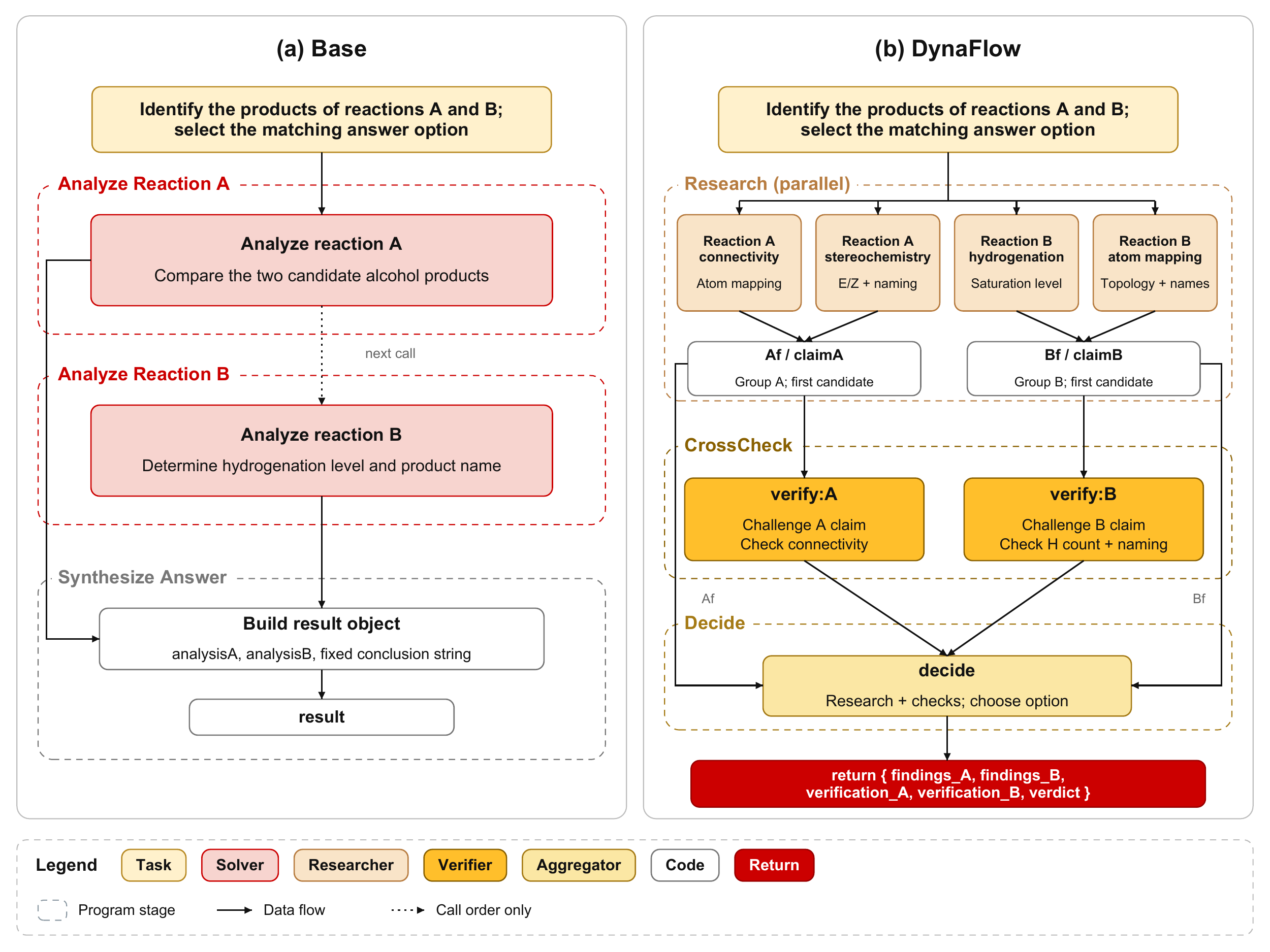}
  {\textbf{GPQA-Diamond: reaction-product identification.} The dotted Base edge denotes call order without a data dependency; the displayed Base program contains no explicit \texttt{return} statement. DynaFlow shows its second invocation, integrating research evidence and verification results before answer selection.}
  {fig:case-gpqa77}

\paragraph{Connecting entity identification to subsequent retrieval.}
For the HotpotQA question requiring person identification and year retrieval (Figure~\ref{fig:case-hotpot521}), the displayed Base call searches for investment firms and their founding years. DynaFlow instead uses three search routes to identify candidate entities, deduplicates and ranks them, and passes their names, quotations, and sources to candidate-specific resolution agents. Subsequent checks assess the selected proposal. This organization connects retrieval steps through explicit intermediate entities and supporting evidence, as discussed in \S\ref{sec:main-results}.

\DynaAppendixFigure
  {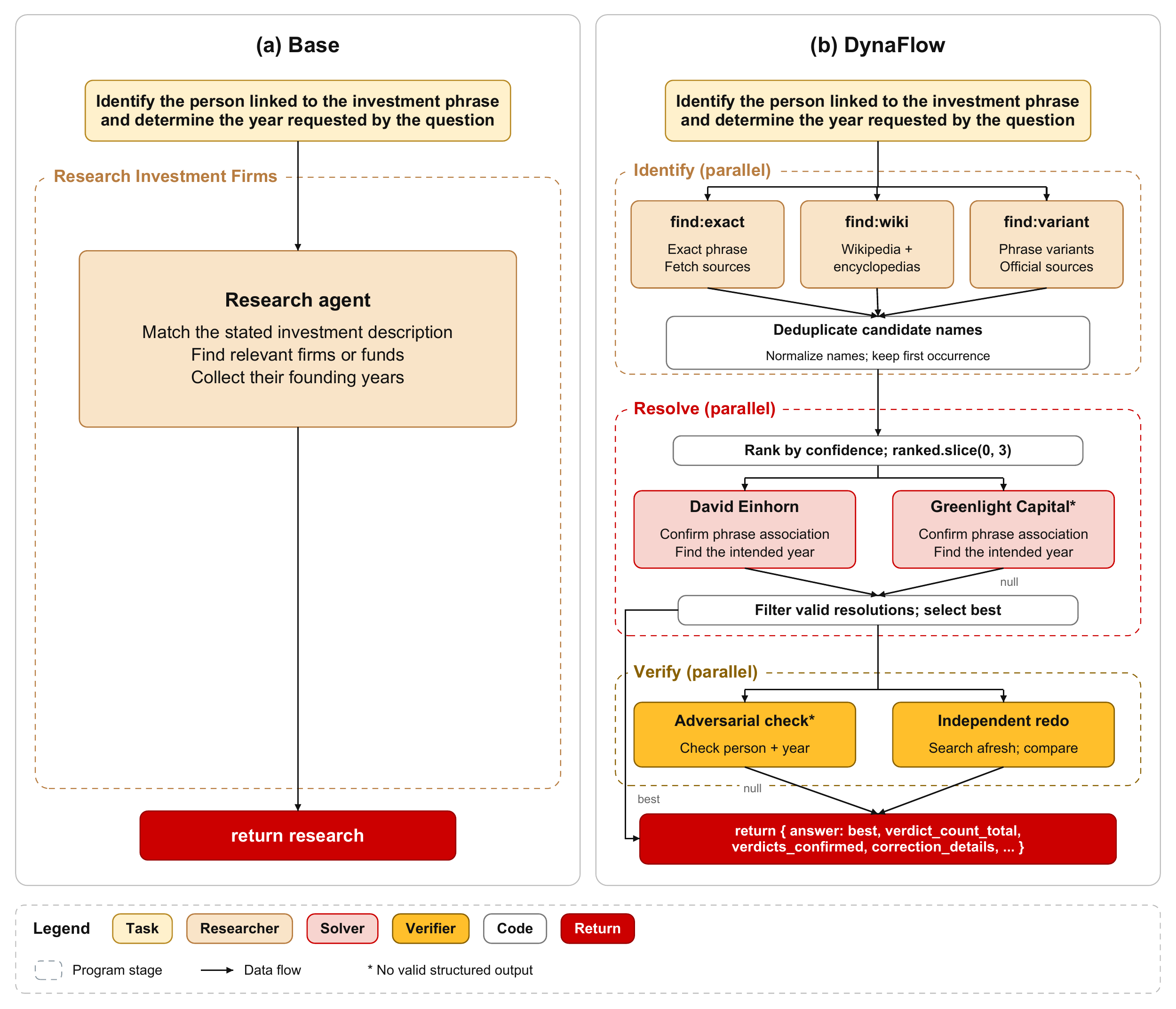}
  {\textbf{HotpotQA: person identification and year retrieval.} Base shows the first of three invocations. DynaFlow resolves two candidates under a top-three cap; starred calls lack valid structured outputs and are filtered out. For person candidates, resolution prompts prioritize birth-year retrieval.}
  {fig:case-hotpot521}

\paragraph{Linking solution stages to audits and conditional repair.}
For the geometric optimization problem from AIME 2025 (Figure~\ref{fig:case-aime242}), the displayed Base invocation checks a supplied derivation and returns fixed numeric fields. DynaFlow connects geometric construction, parallel optimization, expression fitting, and independent audits. Optimizer outputs inform the fitted candidate, while audit findings determine whether to invoke a repair step. Together, these cases illustrate the observation in \S\ref{sec:workflow-analysis}: task-specific orchestration connects stages by specifying what agents receive, how they use it, and how their outputs guide subsequent work.

\DynaAppendixFigure
  {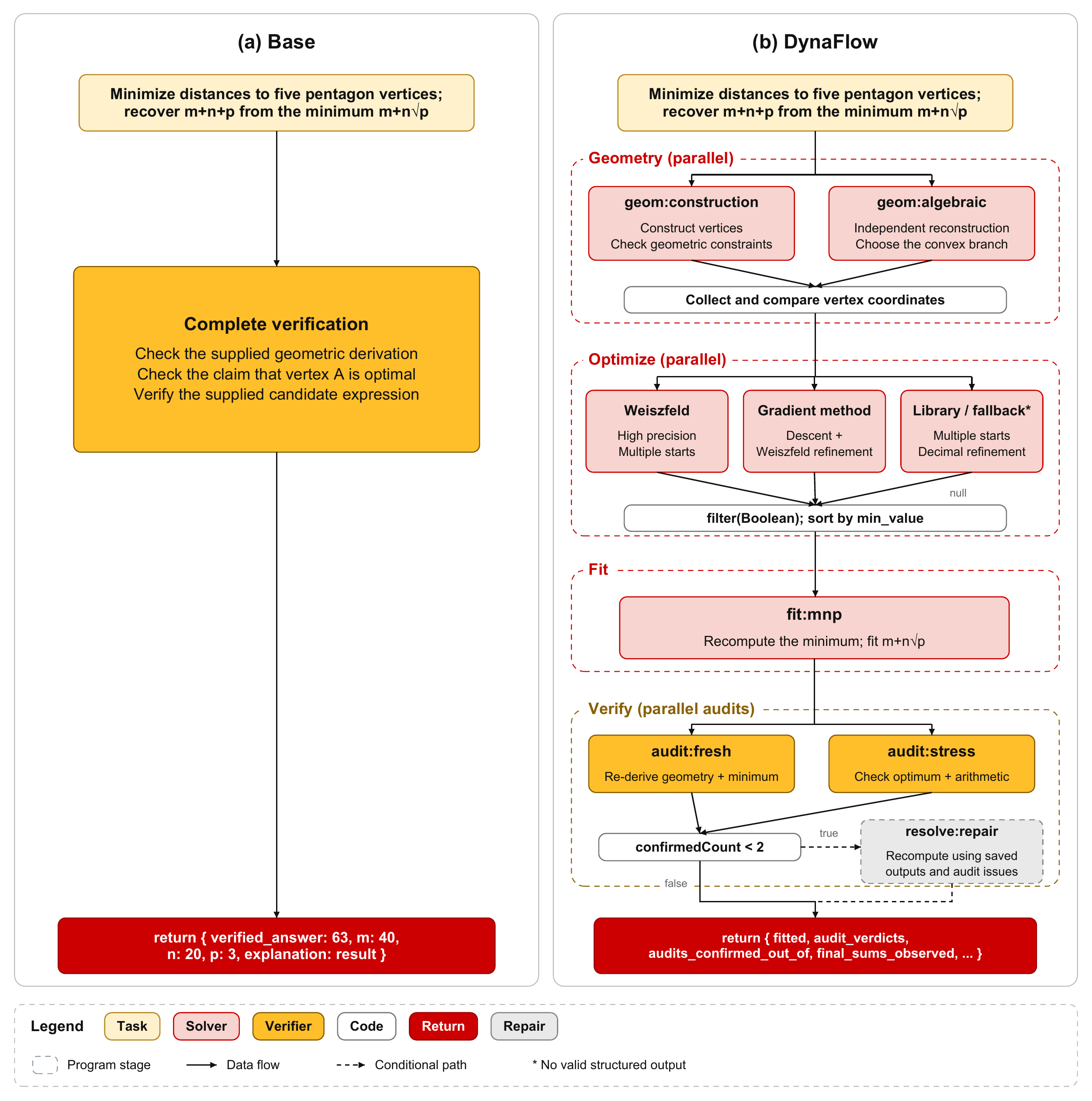}
  {\textbf{AIME 2025: geometric optimization.} Base shows its second invocation; DynaFlow shows its first. The starred optimizer was excluded before fitting because it returned no valid output. Both audits confirmed the candidate, so the dashed conditional repair call was not executed.}
  {fig:case-aime242}

\DynaAppendixClearPage
\section{Extended Related Work}
\label{app:extended-related-work}

\paragraph{Multi-Agent Collaboration and Orchestration.}
LLM-based multi-agent systems (MAS) have progressed from predefined role-based collaboration toward adaptive, task-specific orchestration. Early approaches establish structured interactions through specialized agents, communication protocols, and iterative reasoning~\citep{chen2024agentverse,qian2024chatdev,du2024multiagentdebate,wang2025mixtureofagents}. Recent methods further explore dynamic agent selection, communication topology optimization, and automatic workflow construction~\citep{liu2024dylan,zhang2025gdesigner,li2026assembleyourcrew}.

Similar principles of specialization and selective information integration have been explored in multimodal learning, including cross-spectral augmentation~\citep{lei2024rle}, modality-decoupled representations~\citep{feng2025mdreid}, sparse mixture-of-experts~\citep{feng2025multi}, attention-guided alignment~\citep{tan2026knowing}, and dynamic cross-modal coordination~\citep{lin2026dyco}.

Unlike approaches relying on predetermined collaboration structures, DynaFlow learns to synthesize executable workflows according to individual task requirements and adapt subsequent decisions using execution feedback.

\paragraph{Efficiency and Reliable Information Flow.}
Effective collaboration requires balancing reasoning quality, computational efficiency, and reliable information exchange. AgentPrune reduces redundant communication~\citep{zhang2025agentprune}, while broader MAS efficiency studies investigate topology optimization and resource allocation~\citep{li2026efficiency}. Meanwhile, analyses of multi-agent failures highlight challenges involving task decomposition, information propagation, and verification~\citep{cemri2025failures}.

Related research has explored robust information selection through saliency-guided transfer~\citep{tan2024occluded}, semantic label verification~\citep{zhang2025trustclip}, uncertainty-guided expert selection~\citep{zhang2025choose}, and generalization-preserving or collaborative continual learning~\citep{zhang2025generalization,zhang2026rep}. Reliable evidence integration and structured information dependencies also arise in multimodal hallucination mitigation~\citep{shu2026semantics}, dense video correspondence~\citep{wu2026pairformer}, and constrained visual generation~\citep{wu2025animateqr}.

These studies highlight complementary aspects of information selection, coordination, and reliability. DynaFlow addresses these challenges at the agent-workflow level by selectively invoking collaboration, explicitly organizing dependencies, and training the Main Agent on demonstrations filtered for orchestration quality.

\paragraph{Positioning of DynaFlow.}
DynaFlow unifies task-specific workflow synthesis, adaptive collaboration decisions, and feedback-driven orchestration within a single policy. Unlike methods that optimize fixed workflows or separately train multiple agents, DynaFlow updates only the Main Agent while keeping sub-agents fixed. Its emphasis on explicit information transfer and independent verification enables flexible collaboration tailored to individual tasks.



\end{document}